\documentclass{pkas}

\def\year{2018} 
\def\month{May} 
\def\volume{01} 
\def\issue{0} 
\def\beginpage{1} 
\def\endpage{99} 
\def\received{May 1, 2018} 
\def\accepted{June 27, 2018} 
\date{Received \received ; accepted \accepted}

\title{Recent progress in high-mass star-formation studies with ALMA}

\author[1,2]{Tomoya~Hirota}

\affil[1]{Mizusawa VLBI Observatory, National Astronomical Observatory of Japan, 
Osawa 2-21-1, Mitaka-shi, Tokyo 181-8588, Japan; \email{tomoya.hirota@nao.ac.jp}}
\affil[2]{Department of Astronomical Sciences, SOKENDAI (The Graduate University for Advanced Studies), Osawa 2-21-1, Mitaka-shi, Tokyo 181-8588, Japan}

\def\corrauthor{T. Hirota}

\def\runningauthor{T. Hirota}

\def\runningtitle{High-mass star-formation studies with ALMA}

\def\keywords{stars: formation --- stars: massive --- submillimeter: ISM  --- instrumentation: interferometers}

\def\abstracttext{
Formation processes of high-mass stars have been long-standing issues in astronomy and astrophysics. 
This is mainly because of major difficulties in observational studies such as a smaller number of high-mass young stellar objects (YSOs), larger distances, and more complex structures in young high-mass clusters compared with nearby low-mass isolated star-forming regions (SFRs), and extremely large opacity of interstellar dust except for centimeter to submillimeter wavelengths. 
High resolution and high sensitivity observations with Atacama Large Millimeter/Submillimeter Array (ALMA) at millimeter/submillimeter wavelengths will overcome these observational difficulties even for statistical studies with increasing number of high-mass YSO samples. 
This review will summarize recent progresses in high-mass star-formation studies with ALMA such as clumps and filaments in giant molecular cloud complexes and infrared dark clouds (IRDCs), protostellar disks and outflows in dense cores, chemistry, masers, and accretion bursts in high-mass SFRs. 
}

\begin{document}
\pkashead

\section{Introduction}
\label{sec-introduction}
\quad
High-mass stars defined by the mass of 8$M_{\odot}$ or larger have extremely strong radiation field and stellar wind with high luminosity of $>10^{3}L_{\odot}$, and significantly affect their surrounding environments dynamically and chemically. 
High-mass stars are progenitors of supernova which enrich heavy elements in interstellar matters by nucleosynthesis during stellar evolution and supernova explosion. 
Thus, they have been contributing to cosmic evolution, galaxy formation and evolution, and star-formation in galaxies. 
Despite a large impact on astrophysics and astrochemistry, formation of high-mass stars remains poorly understood due to short evolutionary timescales, clustering, large distances, and heavy obscuration \citep{Zinnecker2007, Tan2014, Motte2018}. 

According to the stellar initial mass function (IMF), which itself is still under debate, higher mass stars are rarely formed compared with lower-mass objects. 
Furthermore, higher mass stars have shorter lifetime of an order of $<$1~Myr in comparison with the Sun with its expected lifetime of 10~Gyr. 
Thus, high-mass stars, in particular for those in early evolutionary phase are extremely rare, and hence, there are little newly born high-mass young stellar objects (YSOs)
or high-mass star-forming regions (SFRs) in the Solar neighborhood. 
Although the nearest low-mass SFRs are located at 120-140~pc from the Sun, even the nearest site of high-mass star-formation, i.e. Orion Molecular Cloud, is located three times larger distance of 420~pc. 
Except for a few examples like Orion, typical distances of high-mass SFRs are larger than 1~kpc and they are distributed in entire regions of our Galaxy. 
The large distances of target sources make it difficult to achieve enough high resolution and sensitivity for detailed studies on their properties. 
The problem of spatial resolution is very serious as high-mass stars are usually formed in dense clusters. 
It prevents us from statistical studies on high-mass star-formation observationally unless each cluster member can be resolved at sufficiently high resolution. 

\begin{figure}
\centering
\includegraphics[width=65mm]{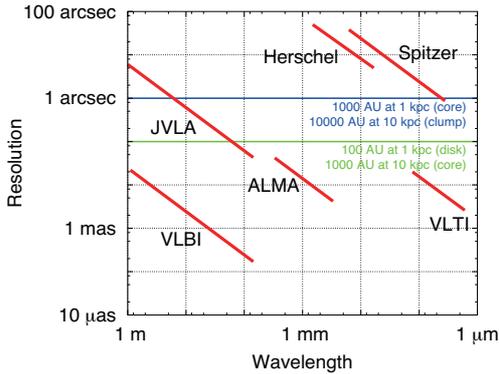}
\caption{Examples for achievable angular resolutions. 
For interferometers (VLBI, JVLA, ALMA and VLTI), the red solid lines indicate the highest resolutions corresponding to the longest baselines. 
Representative angular scales which can be resolved with 0.1 and 1 arcseconds beam sizes are indicated by green and blue horizontal lines, respectively. 
For example, a circumstellar disk with a diameter of 100~au at the source distance of 1~kpc corresponds to the angular size of 0.1 arcsecond (green line). 
Such a structure can be resolved by VLBI, ALMA, and VLTI, but not by JVLA at longer wavelengths than $\sim$1~cm. 
\label{fig-size}}
\end{figure}

The large scale surveys with infrared and submillimeter space instruments starting from IRAS, ISO to Herschel, Planck, Spitzer, AKARI, etc., have contributed to enhance numbers of possible candidates for high-mass SFRs and YSOs deeply embedded in infrared dark clouds (IRDCs). 
They can be observed only in centimeter to far-infrared wavelengths because of large dust opacities of IRDCs even in mid-infrared wavelengths. 
Unfortunately, such large surveys can be done only by satellite telescopes which have less than 1-10" resolution due to limited aperture sizes (Figure \ref{fig-size}). 

As a result, there still remains fundamental questions for high-mass star-formation processes: 
How can high-mass YSOs accrete their mass within short lifetime against strong feedback? 
What are initial conditions for high-mass star-formation?
Why high-mass stars are rarely formed in relation to the stellar initial mass function?
How high-mass binaries and clusters are formed in terms of high-mass star formation processes?
To solve these problems, detailed studies on dynamical and physical properties are necessary for various targets in different environments and evolutionary phases from host clouds to high-mass YSOs. 

In the last decades, newly constructed and renovated observational facilities in various wavelengths have provided high sensitivity and high resolution observations to solve the above fundamental questions. 
Radio interferometer observations from centimeter, millimeter, to submillimeter, such as Jansky Very Large Array (JVLA),  Northern Extended Millimeter Array (NOEMA), Submillimeter Array (SMA), Australia Telescope Compact Array (ATCA), and very long baseline interferometers (VLBI) such as Very Long Baseline Array (VLBA), European VLBI Network (EVN), VLBI Exploration of Radio Astrometry (VERA), Korean VLBI Network (KVN) and VERA Array (KaVA), and Australian Long Baseline Array (LBA), have been playing crucial roles to reveal innermost parts of high-mass star-formation sites deeply embedded in IRDCs. 

In particular, recent progress in Atacama Large Millimeter/Submillimeter Array (ALMA) has improved sensitivities and resolutions by one order of magnitude or larger than previously available connected arrays. 
Furthermore, new capabilities with ALMA have opened millimeter and newly equipped submillimeter windows such as a large instantaneous bandwidth covering numerous molecular lines, wide field mosaic mapping achieving both the higher resolution and larger field of view, high fidelity imaging with multiple 12~m array configurations and Atacama Compact Array (ACA) named as Morita Array, and full polarization observations. 
Figure \ref{fig-size} shows angular resolutions achieved with the recent major astronomical facilities from radio to infrared wavelengths. 
Except for VLBI, the highest resolution achievable from centimeter to submillimeter and near infrared wavelengths are an order of 0.01-0.1" which enables us to resolve the 10-100~AU scales at 1-10~kpc distances. 
Such high resolutions shed light on circumstellar disks and launching regions of outflows close vicinity to newly born high-mass YSOs as demonstrated in low-mass YSOs. 

\section{Summary of ALMA Outcomes}
\quad 
This review summarizes the scientific highlights for high-mass star-formation studies with ALMA obtained from the beginning of the ALMA regular operation in 2011 to the time of the East Asian ALMA science workshop 2017, November. 
It should be noted that only representative observational results from ALMA are presented in this review. 
The readers should refer to original papers for other studies with ALMA as listed in the references and for those other than from ALMA results which are cited in these ALMA papers. 

Since the start of early science operation in 2011, ALMA has completed 5 observing cycles 0-4 until 2017 November, and new cycle 5 session has just started. 
At the time of East Asian ALMA science workshop 2017, more than 800 refereed journal papers have been published using the ALMA data (ESO Telescope Bibliography\footnote{http://telbib.eso.org}), and about 100 of them are related to high-mass star-formation studies. 
Quarter of them ($\sim$23) are produced from the ALMA Science Verification (SV) data on Orion~KL (2011.0.00009.SV), the nearest high-mass SFR \citep[e.g.][]{Zapata2012, Hirota2012, Galvan-Madrid2012, Niederhofer2012}. 
In addition to other individual works \citep[e.g.][]{Hirota2014a, Hirota2014b, Hirota2015, Hirota2016a, Hirota2016b, Hirota2017}, 40 papers are published for Orion KL. 
The other well studied sources are in the Galactic Center region, Sgr B2 \citep[e.g.][]{Belloche2014, Higuchi2015b} and G000.253+0.016 \citep[e.g.][]{Bally2014, Rathborne2014, Higuchi2014}, for which 18 papers are published. 
These Galactic Center sources are located in different environments compared with other typical high-mass star-formation sites in the Galaxy, and hence, they are out of the scope of this review. 
Rest of about 40 papers are for studies on other Galactic high-mass SFRs. 
High-mass SFRs in the Large Magellanic Cloud (LMC) have been studied by ALMA with high sensitivities and resolutions of $<$sub-pc scales \citep[e.g.][]{Fukui2015, Nayak2016, Saigo2017}. 
Physical and dynamical properties of individual SFRs in the LMC can be directly compared with those in the Galactic SFRs, although the star formation in external galaxies in different environments is beyond the scope of the later part of the review. 
The number of high-mass SFR and YSO samples are still limited and further detailed studies, in particular at higher resolutions as discussed later, are strongly desired. 

\begin{table*}[t]
\caption{Typical properties of high-mass SFRs 
\label{tab-property}}
\centering
\begin{tabular}{lcccc}
\toprule
 & Size & Mass & Density & Temperature  \\
\midrule
Cloud                  & $>$1~pc                        & $>$100$M_{\odot}$      & 10$^{2}$-10$^{4}$~cm$^{-3}$  & 10-20~K       \\ 
Clump/filament   & 0.1-1~pc                        & $\sim$100$M_{\odot}$ & 10$^{4}$-10$^{6}$~cm$^{-3}$  & 10-20~K        \\ 
Core (prestellar) & 10$^{3}$-10$^{4}$~AU & $\sim$10$M_{\odot}$    & $>$10$^{6}$~cm$^{-3}$           & 10-20~K        \\ 
Core (HMC)       & 10$^{3}$-10$^{4}$~AU & $\sim$10$M_{\odot}$    & $>$10$^{6}$~cm$^{-3}$           & $\sim$100~K \\ 
Disk                    & 10$^{2}$-10$^{3}$~AU & $\sim$1$M_{\odot}$     & $>$10$^{7}$~cm$^{-3}$           & 100-1000~K  \\ 
\bottomrule
\multicolumn{5}{l}{Note --- The above quantities are different from source to source by a factor of 10. }
\end{tabular}
\end{table*}

\section{Basic Concepts of High-mass Star-formation}
\begin{figure*}[ht]
\centering
\includegraphics[width=180mm, angle=90]{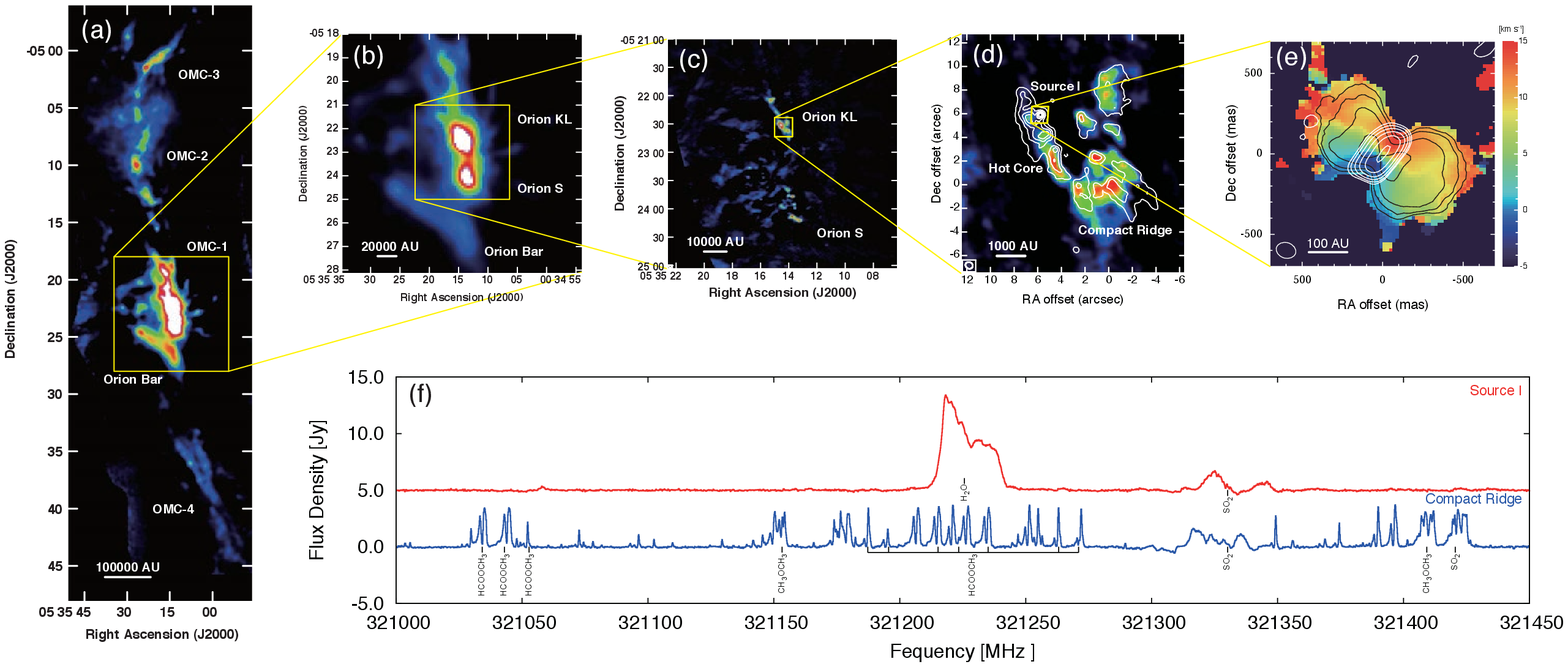}
\caption{Examples of high-mass SFRs in Orion at various scales.
(a) Filament of GMC in Orion A Molecular Cloud observed in the  850 $\mu$m dust continuum emission with JCMT SCUBA \citep[archive data from][]{DiFrancesco2008}. 
(b) Zoom-up of Orion Molecular Cloud 1 (OMC-1) \citep[][]{DiFrancesco2008}. 
(c) Dense cores in OMC-1 observed in the ALMA band 3 continuum \citep[Archive data from ADS/JAO.ALMA\#2015.1.00669.S; see also][]{Hacar2018}. 
(d) Dense cores in Orion KL traced by HCOOCH$_{3}$ line (color, moment 0) and continuum (contour) emission at ALMA band 7 \citep[Archive data from ADS/JAO.ALMA\#2011.0.00199.S; see also][]{Hirota2014b, Hirota2015}. 
(e) Outflow and disk associated with Orion Source~I traced by Si$^{18}$O line (color, moment 1) and continuum (contour) emission at ALMA band 8, respectively \citep[Archive data from ADS/JAO.ALMA\#2011.0.00199.S and 2013.1.00048.S; see also][]{Hirota2016b, Hirota2017}. 
The data for panels (c)-(e) are taken with ALMA while those of (a)-(b) are observed with JCMT (single-dish). 
Extended emissions are significantly resolved out with ALMA, in particular at higher resolution data. 
The most striking differences are seen in panels (b) and (c). 
(f) Examples of molecular spectra in Orion KL Compact Ridge and Source~I \citep[Archive data from ADS/JAO.ALMA\#2011.0.00199.S; see also][]{Hirota2014a, Hirota2014b}.
Representative molecules are indicated by vertical lines. 
Note that the Orion Compact Ridge is also known as rich in molecular species but is distinguished from the Hot Core where oxygen- and nitrogen-bearing organic molecules are abundant, respectively. 
In contrast, Source~I shows only high excitation lines such as H$_{2}$O, SiO, SO$_{2}$, and some maser lines \citep{Plambeck2016}. 
\label{fig-orion}}
\vspace{5mm} 
\end{figure*}

\quad
Here the basic concepts of high-mass star-formation are briefly introduced. 
More details, in particular for theoretical aspects, are referred to other comprehensive review papers \citep{Zinnecker2007, Tan2014, Motte2018}. 
Table \ref{tab-property} summarizes typical properties in different scales of high-mass star-formation sites. 
Figure \ref{fig-orion}(a)-(e) depict relevant structures observed in the nearest high-mass SFR in Orion, although observed properties in Orion cannot always be generalized to typical high-mass SFRs. 

High-mass stars are formed in giant molecular clouds (GMCs) with the hydrogen (H$_{2}$) density and temperature of $\sim$10$^{2}$-10$^{4}$~cm$^{-3}$ and $\sim$10-20~K, respectively. 
Typical sizes range from sub-pc to pc, corresponding to arcseconds or larger angular sizes at distances of the Galactic scale ($\sim$kpc). 
The host clouds are required to have much larger masses than an order of 100$M_{\odot}$ to form multiple high-mass stars and star clusters with $>8M_{\odot}$ for each member. 
Inside the host GMCs, there are smaller scale clumps of higher density molecular gas. 
They are sometimes seen in IRDCs which can be recognized as absorption in near- and mid-infrared wavelengths. 
Many IRDCs show characteristic filamentary elongated structures with smaller fragmentations.  
Inside these clumps, high-mass young stellar objects (YSOs) are thought to be formed in the central part of compact dense cores. 
In general, high-mass stars are formed in binaries or a small number of multiple systems inside the cores. 
After the onset of high-mass star-formation, dense cores are heated internally via stellar radiation and/or outflow shocks. 
As a result, they can be identified as hot cores at a higher temperature of $\sim$100~K. 
These cores have compact structures with 10$^{4}$~AU or smaller and hence, high resolutions of 1" or better are required to spatially resolve their structures. 
Around the newly born YSOs, there are circumstellar disks and outflows as seen in low-mass YSOs. 
Outflows are extended from 100 to 10$^{4}$~AU scales depending on their evolutionary phases, but disks are always compact with 100-1000~AU scales or 0.1"-1" even at the distance of 1~kpc. 
Because high-mass YSOs are usually formed in dense star clusters, they have significant feedback to their surrounding media due to their strong radiation field, stellar winds, and outflows, as seen in extended optical/infrared nebular and H{\sc{ii}} regions. 

\section{Filaments, Clumps, and Cores}
\quad
High-mass YSOs are formed in massive reservoirs of accreting materials. 
Currently, there are two controversial major scenarios for high-mass star-formation. 
One is turbulent core accretion \citep{McKee2002} and another is competitive accretion \citep{Bonnell2001}, and there are several modified/advanced theories. 
The turbulent core accretion model is a so-called scaled-up version of low-mass star-formation theories and is also referred to as a monolithic collapse model in which a YSO (or binary or a small number of multiples) is formed by collapse of a single dense core. 
Because the thermal Jeans mass is too small to form high-mass stars, the thermal support is insufficient for dense cores potentially forming high-mass stars. 
For this, it is predicted that high-mass starless cores are supported via turbulence or alternatively magnetic fields to form a single high-mass star (or a binary or a small number of multiples) in each core. 
The initial condition in the cores is thought to be close to the virial equillibrium in the turbulent core accretion model. 
On the other hand, the competitive accretion model predicts a mass assembly through global gravitational forces in the central part of the clumps surrounded by smaller scale multiple cores. 
In contrast to the turbulent core accretion model, it is predicted that there are only low-mass fragments or cores with the thermal Jeans mass close to the possible site of high-mass star-formation. 
Each high-mass star increases its mass via Bondi-Hoyle accretion at the center of cluster formation sites. 
The competitive accretion model predicts that the cores are in sub-virial state and show rapid global collapse. 

To investigate initial conditions prior to high-mass star-formation, IRDCs are recognized as ideal laboratories because they are in a deeply embedded prestellar phase under gravitational collapse or mass accretion. 
One promising approach to distinguish two high-mass star-formation theories is to search for high-mass starless cores which are predicted only for the turbulent core accretion model.  
Prototypical examples are IRDC cores G11.92-0.61-MM2 identified by observations with SMA and JVLA \citep{Cyganowski2014}, C1-N, C1-S, and C9A in G028.37+00.07 detected with ALMA \citep{Tan2013, Tan2016, Kong2016, Kong2017}. 
In case of in G028.37+00.07, two dense cores, C1-N and C1-S, were first identified as possible candidates of very young evolutionary phase of high-mass star-formation because of no sign of star-formation activity and high deuterium fractionation in the N$_{2}$D$^{+}$ $J$=3-2 line, which is a possible indicator of chemically evolved phase and slow gravitational collapse timescale \citep{Tan2013}. 
They have masses of $\sim$60$M_{\odot}$ and are though to be magnetically virialized which can form a high-mass YSO in each core as predicted via the turbulent core accretion model. 
Subsequent follow-up observations detected molecular outflows from one of the cores, C1-S, \citep{Tan2016, Kong2016, Feng2016}, suggesting that it is a deeply embedded high-mass YSO. 
Although another one, C1-N, could be a starless prestellar core candidate, there are little other convincing evidences for such high-mass starless cores, and hence, further searches are required. 

At pc-scale structures, IRDCs are known to have filamentary structures showing possible signatures of global accretion motions. 
Many observational studies have been done with ALMA targeting at filaments, clumps, and cores in IRDCs \citep{Sakai2013, Sakai2015, Liu2015, Ohashi2016, Minh2016, Fontani2016, Maud2017, Henshaw2017}. 
ALMA can provide detailed spatial and velocity structures in individual filaments. 
For instance, an IRDC clump SDC335.579-0.272 (SDC335) shows a hub-filaments structure in the N$_{2}$H$^{+}$ $J$=1-0 line \citep{Peretto2013, Avison2015}. 
SDC335 is thought to be the most massive core in the Galaxy with a mass of 545$M_{\odot}$ (MM1) at a mass flow rate of 2.5$\times 10^{-3}M_{\odot}$yr$^{-1}$ along the filaments \citep{Peretto2013, Avison2015}.  
The high mass accretion and large mass reservoir suggest that SDC335 is a potential site of OB cluster formation in the central core around the hub-filaments structure via the global collapse. 

The sequence of star-formation in each high-mass SFR is also a matter of debate for understanding of high-mass star-formation processes.  
Two high-mass star-formation theories result in different mass accretion processes and consequently different history of star-formation. 
In the turbulent core accretion model, a high-mass YSO (or multiple system) can be formed in a single massive core. 
On the other hand, the competitive accretion model expects high-mass star-formation in central parts of the cluster formation surrounded by low-mass cores and YSOs. 
High sensitivity ALMA imaging can detect substellar-mass cores even in distant high-mass star/cluster-forming regions. 
These observations are useful to test above differences by identifying potential sites of low-mass star-formation therein. 
To date, there have been two different observational results favoring (or being consistent with) the two controversial models. 
Extreme cases are high-mass IRDCs G28.34+0.06 P1 \citep{Zhang2015} and G11.92-0.61 \citep{Cyganowski2017}. 
In the former case for G28.34+0.06 P1, a high-mass filament with 10$^{3}M_{\odot}$ has fragments of 5 cores with a mass of 20-43$M_{\odot}$ for each, while there is no low-mass population distributed around a high-mass filament with the upper limit of 0.2$M_{\odot}$, which is 30 times lower than the thermal Jeans mass \citep{Zhang2015}. 
Thus, it is thought that low-mass stars will be formed later than higher mass cluster members in G28.34+0.06 P1. 
On the other hand, high-mass cores in G11.92-0.61 ($>$30$M_{\odot}$) are surrounded by several low-mass cores and YSOs with $\sim$1$M_{\odot}$ \citep{Cyganowski2017}. 
The low-mass cores in G11.92-0.61 show clear signatures of molecular outflows, suggesting that both high- and low-mass YSOs are forming simultaneously, as predicted by the competitive accretion model. 
Combined with other observational studies on protoclusters \citep{Brogan2016, Foster2014, Liu2017}, more statistical datasets would be necessary to judge which is the more dominant processes in high-mass star-formation. 

\section{Disks and Outflows}
\quad 
One of the well known problems in high-mass star-formation theory is a radiation feedback from newly born YSOs. 
Because of the shorter Kelvin-Helmholtz timescale than that of accretion, high-mass stars with mass of $\sim$10$M_{\odot}$ become zero-age main-sequence still under accretion phase to grow up higher-mass stars. 
However, the strong radiation pressure working on the dusty envelope around newly formed YSOs would halt mass accretion against their gravity in the case of a spherically symmetric accretion structure \citep{Wolfire1987}. 
The solution for this problem is to introduce a high-mass accretion rate of the order $>$10$^{-3}M_{\odot}$~yr$^{-1}$ and/or a non-isotropic accretion geometry. 
If the surrounding envelopes form rotating disks around newly born YSOs due to their angular momentum, the strong radiation can preferentially escape from the polar direction through the outflow cavity and consequently, the mass accretion can be continued through disks, which is known as a flashlight effect \citep{Yorke1999}. 

Even before the ALMA era, clear signatures of rotating and/or accretion disks around high-mass B-type YSOs have been reported from observational results of millimeter interferometers, near-infrared interferometry and VLBIs \citep{Cesaroni2007, Beltran2016a}. 
Thus, it has been accepted that high-mass B-type stars can be formed through disk accretion similar to low-mass YSOs, while the number of well studied samples with high resolution is still limited. 
In contrast, it is unclear whether more massive O-type stars can be formed in a similar way because such objects usually show more massive ($>100M_{\odot}$) and larger ($>10^{4}$~AU) rotating structures called toroids \citep{Cesaroni2007, Beltran2016a}. 

The higher resolution and higher sensitivity ALMA data provide more samples with detailed velocity structures showing rotation motions \citep{Sanchez-Monge2013, Sanchez-Monge2014, Beltran2014, Guzman2014, Hirota2014a,  Zapata2015} even in O-type stars suggesting the same formation processes \citep{Beltran2016a}. 
One of the best examples for the Keplerian-like rotation in which the rotation velocity is inversely proportional to the square root of the radius, $v_{\rm{rot}} \propto r^{-0.5}$, is an O7-type (25$M_{\odot}$) high-mass YSO AFGL4176 \citep{Johnston2015}. 
They utilize multi-transitions of the CH$_{3}$CN lines in ALMA Band 6 at different excitation energy levels, which trace the different temperature regions and hence, different radius, to construct a model of the Keplerian rotation disk with the 2000~AU radius. 
Another study of an O-type YSO, G351.77-0.54, also shows a 1000~AU scale rotating disk in the CH$_{3}$CN lines at a resolution of 130~AU \citep{Beuther2017}. 
The absorption lines at higher-frequency in ALMA Band 9 reveal the infalling motion onto the disk at a high mass accretion rate of 10$^{-4}$-10$^{-3}M_{\odot}$~yr$^{-1}$. 
The highest resolution data shows the temperature distribution in the disk suggesting that it is gravitationally stable against fragmentation based on the discussion on the Toomre Q-parameter. 
Although a pilot survey of disks around O-type stars shows more disturbed and complicated structures \citep{Cesaroni2017}, further survey data at high resolutions will provide more detailed dynamical properties for understanding high-mass YSOs and binaries formation. 

Outflows are one of the most outstanding phenomena in star-formation activities as they are more extended than those of the disks. 
Outflows are closely related to mass accretion processes and hence, physically connected to the disks. 
ALMA provides outflow samples in high-mass SFRs \citep{Merello2013, Higuchi2015a, Feng2016, Beuther2017} with their size scales ranging from an order of 100-10000~AU. 
These studies would give their driving sources and mechanisms, (indirectly) mass accretion and feedback process which can regulate star-formation activities. 

Very recently, ALMA has demonstrated definite evidences of rotation motions of outflows and jets for low-mass YSOs \citep{Bjerkeli2016, Lee2017}. 
This can also be achieved for the nearest high-mass YSO, a radio source~I (Source~I) in Orion~KL \citep{Hirota2017}. 
Figure \ref{fig-orion}(e) shows the distribution of the Si$^{18}$O emission at 484~GHz in ALMA Band 8. 
The Si$^{18}$O line shows a consistent velocity gradient with those of more compact high excitation H$_{2}$O lines and SiO masers emitted from the disk \citep{Hirota2014a, Hirota2017}.  
The position-velocity diagram suggests an enclosed mass of 8.7$\pm$0.6$M_{\odot}$ and centrifugal radii of 21-47~AU. 
Along with the expansion velocity estimated to be 10~km~s$^{-1}$ without high-velocity collimated optical or radio jets, the driving mechanism of this low-velocity outflow is most likely explained by a magneto-centrifugal disk wind model \citep{Blandford1982, Matsushita2017}. 
It is likely that disk/outflow system around Orion Source~I could have a similar formation scenario analogous to low-mass YSOs. 

In addition to the above bipolar outflow samples, more energetic explosive outflows are identified. 
Again, the nearest high-mass SFR Orion KL is one of such rare cases \citep{Bally2017}. 
The explosive outflow can be traced by near-infrared and millimeter/submillimeter molecular lines ejecting jet-like streamers almost  isotropically at the velocity of $>$100~km~s$^{-1}$. 
The dynamical timescale of $\sim$500~yrs and  kinetic energy of 10$^{48}$~erg are consistent with the idea that high-mass YSOs in Orion~KL, Source~I and BN, experienced a dynamical interaction about 500~yrs ago to form massive ($\sim$20$M_{\odot}$) close binary system in Source~I \citep{Rodriguez2017}. 
However, this scenario is still controversial because of apparent inconsistency with the lower-mass estimated from the other ALMA observations of the rotating disk in Source~I \citep{Hirota2014a, Hirota2017, Plambeck2016}.\footnote{After the East Asian ALMA science workshop 2017, the higher resolution ALMA observations is reported by \citet{Ginsburg2018}, in which the mass of Source~I becomes more consistent estimated to be 15$\pm$2$M_{\odot}$. }

Although both of the two possible accretion scenarios of high-mass star-formation (i.e. the turbulent core accretion and competitive accretion models) expect the circumstellar disk/outflow systems, they predict different dynamical properties \citep{Tan2014, Beltran2016a}.  
Because the competitive accretion occurs in the central part of dense clusters, disks around newly born high-mass YSOs would be perturbed dynamically, which results in truncation of the smaller disk size and more chaotic directions of outflows. 
Thus, the statistical studies of disk/outflow systems with higher resolutions are also the key to understanding of high-mass star-formation scenarios. 

\section{Chemistry}
\quad 
Once high-mass YSOs are formed in dense cores, they heat surrounding media via radiation and/or outflow shocks up to the sublimation temperature of molecules freezed-out onto grain surface. 
As a result, high-mass cores hosting YSOs just after formation show rich molecular lines in particular from complex organic molecules formed via grain-surface reactions. 
These objects are known as Hot Molecular Core (HMC) or Hot Core (HC). 
Sample spectra for such sources in Orion~KL are shown in Figure \ref{fig-orion}(f). 
The prototypical HMCs are the Hot Core and Compact Ridge in Orion KL as observed in the ALMA SV \citep{Fortman2012} and some other studies \citep[e.g.][]{Plambeck2016,Wright2017}, although recent observations suggest that Orion Hot Core could be externally heated by outflow shocks rather than by the central YSO \citep{Wright2017,Orozco-Aguilera2017}. 

ALMA has opened new windows for high-resolution molecular line maps other than in Orion KL. 
New ALMA data have revealed chemical differentiation among high-mass SFRs and even within each region \citep{Sanchez-Monge2014, Beltran2014, Zhang2015, Watanabe2017}. 
These results show completely different spectral features such that one core shows a rich chemistry while another nearby source with similar physical properties are poor or absent of any molecular lines. 
Such a striking diversity would be due to intrinsic chemical differentiation affected by a mass and luminosity of the heating source, or due to a high opacity of the dust continuum which results in absorption of molecular lines. 
Although the interpretation is still not convincing, chemistry could play key diagnostics of physical properties and their evolution in high-mass star-formation processes. 

Thanks to the high sensitivity, large instantaneous bandwidth, and wide ranges of submillimeter bands, searches for new molecular/atomic lines have been done with ALMA for various kind of sources including low- and high-mass YSOs, late-type stars, and distant galaxies. 
As for high-mass SFRs, ALMA has discovered three new interstellar molecules; isopropyl cyanide (C$_{3}$H$_{7}$CN) in Sgr B2(N) \citep{Belloche2014}, trans ethyl methyl ether (t-C$_{2}$H$_{5}$OCH$_{3}$) in Orion KL \citep{Tercero2015}, and N-methylformamide (CH$_{3}$NHCHO) in Sgr B2(N) \citep{Belloche2017}. 
Search for more complex organic molecules will provide a complete picture of chemistry in high-mass SFRs and possible link to astrobiology (i.e. the origin of terrestrial life). 

\section{Masers and Accretion Bursts}
\quad
Molecular maser emissions such as H$_{2}$O and CH$_{3}$OH are known to be associated with a large number of high-mass YSOs. 
They are predominantly distributed in hot ($>$100-1000~K) molecular gas heated via shocks and/or strong radiation field, and hence, useful tracers for disks and outflows in close vicinity to high-mass YSOs at 100~AU scales. 
Almost all previous observations before the ALMA era were limited to centimeter to millimeter wavelengths such as methanol masers at 6.7~GHz (class II) and 44~GHz (class I), water masers at 22~GHz, and SiO masers at 43~GHz. 
With ALMA, various kind of millimeter and submillimeter masers are predicted \citep{Humphreys2007, Voronkov2012, Perez-Sanchez2013}. 
The submillimeter H$_{2}$O maser lines at 321~GHz (see Figure \ref{fig-orion}(f)) and 658~GHz are detected toward Orion Source~I \citep{Hirota2014a, Hirota2016a}. 
Millimeter and submillimeter CH$_{3}$OH masers are also mapped at 278~GHz in IRDC~G34.43+00.24~MM3 \citep{Yanagida2014} and at 349~GHz in S255IR~NIRS3 \citep{Zinchenko2017}. 
High-sensitivity observations with ALMA have confirmed new SiO maser sources and lines associated with high-mass YSOs \citep{Niederhofer2012, Higuchi2015b, Cho2016}, which are quite rare cases with merely three detection until ALMA. 
These masers are compact and unresolved even with the highest resolution ALMA observations. 
They will be good probes for disks and outflows to achieve the highest resolution observations at 100~AU scales comparable with currently available VLBI technique at the maximum resolution of 1~mas (Table \ref{fig-size}). 

Masers are known to be variable at short ($\sim$days to year) timescales. 
Regarding such variability, flare-like events nearly synchronized to the infrared and radio continuum emissions are reported for high-mass YSOs in NGC6334I-MM1 \citep{Hunter2017} and S255IR~NIRS3 \citep{CarattioGaratti2017}. 
In the case of NGC6334I-MM1, the luminosity is estimated to increase by a factor of 70 (up to 4.2$\times$10$^{4}L_{\odot}$) than that in pre-flare phase \citep{Brogan2016, Hunter2017}. 
This can be explained by a sudden increase in a mass accretion rate as proposed for an accretion burst in low-mass FU-Ori and EX-Lup objects. 
Thus, such maser flares are recognized as one of key processes to understand the disk-mediated episodic accretion or accretion burst in star-formation processes. 

Burst- or flare-like activities in maser emission have also been recognized for more than 30~years since the discovery of "Super-maser" in Orion KL and W49N. 
The H$_{2}$O maser burst occurred in 2011, just before the start of ALMA Early Science cycle 0 \citep{Hirota2014b}. 
Although no submilliemter H$_{2}$O line shows such a flare activity unlike that of the 22~GHz H$_{2}$O maser, the data could constrain the emission mechanisms and physical properties of the supermaser region. 
Further monitoring observations in the ToO mode with ALMA are crucial for coordinated observations of time-domain studies on mass accretion in both high- and low-mass YSOs. 

\section{Future Prospects}
\quad 
For future high-mass star-formation studies with ALMA, there are several directions; higher resolution, higher sensitivity, larger field of view, wider frequency coverage, and other new capabilities such as polarization and astrometry. 
ALMA has not yet achieved the highest resolution at the most extended configuration except for lower-frequency than Band 6. 
Thus, there still remains a room to improve resolution better than 10~mas. 
It is almost comparable to those achieved by VLBI arrays at lower frequencies ($\sim$a few GHz) (Figure \ref{fig-size}). 
Thus, the ALMA data can be combined with the VLBI mapping and proper motion measurements of maser lines to reveal three-dimensional velocity fields \citep[e.g.][]{Beltran2016b}.  
It should be noted that the brightness sensitivity is inversely proportional to the square of the beam size. 
Although the continuum emission will be detectable at sufficiently high sensitivity thanks to the large bandwidths of ALMA, the decrease in brightness sensitivity will be serious for thermal molecular line observations at high resolution configurations. 
Some of the bright millimeter/submillimeter maser lines will be potential probes for the highest resolution observations as mentioned in the previous section. 

Another more important issue is a capability of polarization measurements. 
Polarized emissions provide information on magnetic fields toward the target sources. 
The magnetic field is thought to play a crucial role in star-formation processes because it regulates formation of filamentary structures in IRDCs, contraction of dense cores, mass accretion onto YSOs, and formation of disks and outflows \citep{Crutcher2012}. 
For this purpose, the linear polarization of the dust continuum emission and maser lines \citep{Perez-Sanchez2013} will provide the magnetic field direction from disks and dense cores to larger scale filamentary structures through high dynamic range polarization images with ALMA. 
The circular polarization of molecular lines, especially radicals with unpaired electrons (SO, CCS, CN) and strong maser species (H$_{2}$O, CH$_{3}$OH, SiO) can be used for estimating the line-of-sight magnetic field strength through the Zeeman splitting measurements. 
With ALMA, there have been only a few examples for such magnetic field measurements \citep[e.g.][]{Cortes2016}, and more samples are required. 

Because of limited available time and capabilities, high-mass star-formation studies with ALMA have been carried out as compilation of detailed case studies and a small number of pilot surveys such as targeted for 6 O-type YSOs with disks \citep{Cesaroni2017}, 32 high-mass starless core candidates \citep{Kong2017}, 35 massive protoclusters \citep{Csengeri2017}, and 46 ATLASGAL clumps \citep{Chibueze2017}. 
Further statistics with uniform datasets are desired (e.g. ALMA large program; 2017.1.01355.L). 

\section{Summary}
\quad 
ALMA has opened a new window for high-mass star-formation studies with the unprecedented high sensitivity and resolution. 
A number of high quality images of filaments, clumps, and cores in high-mass SFRs provide hints for understanding initial conditions and formation scenarios of high-mass YSOs, either the turbulent core accretion or  the competitive accretion. 
The high resolution achieved with ALMA can resolve the 100-1000~AU scales of disks and outflows associated with newly born high-mass YSOs. 
These results show strong evidences of the accretion through disks and magnetically driven outflows which are analogous to low-mass star-formation. 
Molecular chemistry reveals diversity of high-mass SFRs other than a classical hot core in Orion KL, and demonstrats a potential for discoveries of new molecular species. 

Nevertheless, observations are still limited for case studies of well know sources, and spatial resolutions are insufficient to resolve the innermost regions close to the newly born YSOs at 100~AU or smaller. 
Line-survey observations will unveil a complete picture of molecular composition of newly born HMCs, in particular for complex organic molecules related to astrobiology. 
Coordinated time domain observations with the continuum and maser emissions will provide episodic accretion in high-mass stars. 
More importantly, polarization observations are also limited, but the full polarization capability will be available with ALMA for the magnetic field measurements by the dust continuum emissions and masers in linear polarization and Zeeman splitting measurements of molecular lines in circular polarization. 
Further statistical surveys of a large number of high-mass SFRs and wide-field mappings are crucial for understanding of high-mass star-formation. 

\acknowledgments
\quad
I would like to thank the anonymous referees for critical reading of the manuscript. 
I am also grateful to Masahiro N. Machida and Kazuhito Motogi for valuable discussion on this paper. 
This paper makes use of the following ALMA archive data: ADS/JAO.ALMA\#2011.0.00009.SV, 2011.0.00199.S, 2013.1.00048.S, and 2015.1.00669.S. 
ALMA is a partnership of ESO (representing its member states), NSF (USA) and NINS (Japan), together with NRC (Canada), NSC and ASIAA (Taiwan), and KASI (Republic of Korea), in cooperation with the Republic of Chile. The Joint ALMA Observatory is operated by ESO, AUI/NRAO and NAOJ. 
The author is financially supported by the MEXT/JSPS KAKENHI Grant Numbers 21224002, 24684011, 25108005, 15H03646, 16K05293, and 17K05398.

\end{document}